\documentclass[10pt,aps,rmp,onecolumn,superscriptaddress,floatfix,nofootinbib,showkeys,notitlepage,preprintnumbers]{revtex4-1}
\usepackage{amsmath,amssymb,amsfonts}
\usepackage{graphicx,color}
\usepackage{natbib}
\usepackage{eurosym}
\usepackage{hyperref}
\usepackage[inline]{enumitem}
\usepackage{soul}
\newcommand{\ie}{\emph{i.e.}}
\newcommand{\eg}{\emph{e.g.}}

\begin{document}

\title{What do central counterparties default funds really cover? A network-based stress test answer}

\author{Giulia Poce}
\affiliation{\small Dipartimento di Fisica Universit\`a ``Sapienza'' - 00185 Rome (Italy)}
\author{Giulio Cimini}\email{giulio.cimini@gmail.com}
\affiliation{\small IMT School for Advanced Studies - 55100 Lucca (Italy)}
\affiliation{\small Istituto dei Sistemi Complessi (ISC)-CNR UoS ``Sapienza'' - 00185 Rome (Italy)}
\author{Andrea Gabrielli}
\affiliation{\small IMT School for Advanced Studies - 55100 Lucca (Italy)}
\affiliation{\small Istituto dei Sistemi Complessi (ISC)-CNR UoS ``Sapienza'' - 00185 Rome (Italy)}
\affiliation{\small INFN Roma1 unit - 00185 Rome (Italy)}
\author{Andrea Zaccaria}
\affiliation{\small Istituto dei Sistemi Complessi (ISC)-CNR UoS ``Sapienza'' - 00185 Rome (Italy)}
\author{Giuditta Baldacci}
\thanks{The information and views set out in this paper are those of the authors and do not reflect the opinion of CC\&G S.p.A., London Stock Exchange Group.}
\author{Marco Polito}
\thanks{The information and views set out in this paper are those of the authors and do not reflect the opinion of CC\&G S.p.A., London Stock Exchange Group.}
\author{Mariangela Rizzo}
\thanks{The information and views set out in this paper are those of the authors and do not reflect the opinion of CC\&G S.p.A., London Stock Exchange Group.}
\author{Silvia Sabatini}
\thanks{The information and views set out in this paper are those of the authors and do not reflect the opinion of CC\&G S.p.A., London Stock Exchange Group.}
\affiliation{\small CC\&G S.p.A. London Stock Exchange Group - 00186 Rome (Italy)}
%\date{\today}

\begin{abstract}
In the last years, increasing efforts have been put into the development of effective stress tests to quantify the resilience of financial institutions. 
Here we propose a stress test methodology for central counterparties based on a network characterization of clearing members, whose links correspond to direct credits and debits. 
This network constitutes the ground for the propagation of financial distress: equity losses caused by an initial shock with both {\em exogenous} and {\em endogenous} components 
reverberate within the network and are amplified through {\em credit} and {\em liquidity} contagion channels. 
At the end of the dynamics, we determine the {\em vulnerability} of each clearing member, which represents its potential equity loss. 
We apply the proposed framework to the Fixed Income asset class of CC\&G, the central counterparty operating in Italy whose main cleared securities are Italian Government Bonds.
We consider two different scenarios: a distributed, plausible initial shock, as well as a shock corresponding to the {\em cover 2} regulatory requirement 
(\ie, the simultaneous default of the two most exposed clearing members). 
Although the two situations lead to similar results after an unlimited reverberation of shocks on the network, 
the distress propagation is much more hasty in the latter case, with a large number of additional defaults triggered at early stages of the dynamics. 
Our results thus show that setting a default fund to cover insolvencies only on a {\em cover 2} basis may not be adequate for taming systemic events, 
and only very conservative default funds---such as CC\&G's one---can face total losses due to the shock propagation. 
Overall, our network-based stress test represents a refined tool for calibrating default fund amounts.
\end{abstract} 

\keywords{Stress test; Systemic risk; Financial networks; Central counterparty.}

\maketitle

\section{Introduction}\label{intro}

The financial crises of the last decade have revealed the inherent structural fragilities of the financial system. In this context, the scientific community put substantial efforts 
in understanding the complex patterns of interconnections characterizing financial markets \citep{Boss2004,Iori2006,Nier2008,Krause2012,Battiston2016} and how financial distress 
spreads among financial institutions through direct exposures to bilateral contracts and indirect exposures through common assets ownership \citep{Allen2000,Gai2011,Caccioli2014,Acemoglu2015,Gualdi2016X}. 
Various techniques have thus been developed to study how {\em local} events may trigger a {\em global} instability through amplification effects like default cascades 
\citep{Gai2010,Haldane2011,Corsi2013,Bardoscia2016X}, and to quantify the resulting systemic risk in capital markets 
\citep{Eisenberg2001,Furfine2003,Cifuentes2005,Elsinger2006,Brunnermeier2009a,Lau2009,Battiston2012,Greenwood2015,Glasserman2015}. At the same time, 
regulators were pushed to introduce more stringent rules on capital and liquidity requirements \footnote{{\em Basel III: A global regulatory framework for more resilient banks 
and banking systems}. Basel Committee on Banking Supervision (2010). \url{http://www.bis.org/publ/bcbs189.pdf}} which, coupled with the adoption of micro-prudential policies by commercial and investment banks, 
the promotion of central counterparties (CCPs) as contract intermediaries, and the quantitative easing monetary policy by central banks, eventually increased the robustness of the financial system. 

In this work we focus on CCPs, corporate entities that guarantee the terms of a trade between two {\em clearing members} (CMs)---\ie, financial institutions 
participating in the market cleared by the CCP---in case of insolvency of one of the parties. 
This is achieved by collecting guarantees from each CM for covering potential losses stemming from a missed fulfillment of their clearing obligations, 
resulting in the CCP replacing the trade at the current market price. In particular, a CCP collects two different types of guarantees from its CMs: 
{\em margins} and {\em default fund} amounts. Margins are called on a daily basis to cover the theoretical liquidation costs that the CCP would incur in the event of default of a CM, 
in order to close the open positions under severe market scenarios. 
The default fund instead is a mutualized guarantee fund that aims at covering market risks above and beyond those covered by margins, under the assumption of default of one or more CMs.
The default fund is calibrated through regularly performed stress tests. 
In particular, according to EMIR (the European Market Infrastructures Regulation) 
\footnote{EMIR Regulation (EU) 648/2012 on OTC derivatives, central counterparties and trade repositories, \url{http://data.europa.eu/eli/reg/2012/648/oj}}, these stress tests should determine 
whether the CCP has sufficient resources to cover losses resulting from the default of at least the two CMs to which it has the largest exposure under ``extreme but plausible'' market conditions, 
the so-called {\em cover 2} requirement. 

Given the significant role CCPs play in the stability of the European financial system, the European Securities and Markets Authority (ESMA) has recently coordinated the first EU-wide assessment 
of the resilience of CCPs to adverse market developments, putting a specific focus on the interconnections between the numerous participants in the EU financial system \footnote{ESMA report 
on EU-wide CCP Stress test 2015, \url{https://www.esma.europa.eu/sites/default/files/library/2016-658_ccp_stress_test_report_2015.pdf}, published April 2016}. 
This stress exercise involved 17 CCPs, and focused on the counterparty credit risk that CCPs would face as a result of multiple CMs defaults and simultaneous market price shocks. 
The assessment also included an analysis of the potential spill-over effects to non-defaulting CMs. Indeed, CCPs manage defaults by means of the so-called ``default waterfall'', 
defined in article 45 of EMIR. According to this mechanism, in case of a default all the guarantees posted by the defaulting member (both margins and contribution to the default fund) 
is used first in order to cover potential losses the CCP is facing. However, in case of severe losses this may not be enough, so that a dedicated amount of the CCP's capital 
\footnote{This amount is significantly lower than the default fund, and serves as an incentive for CCPs to collect prudential margin amounts.} is used and, 
at last, the default fund of the non-defaulting CMs is used too, resulting in spillover losses for non-defaulting CMs. Additionally, losses may also derive from the fact that the various CCPs 
are highly interconnected through common CMs, so that the default of one of the top members or groups of a CCP could potentially impact other CCPs as well. 

The ESMA exercise acknowledged that the system of EU CCPs is overall resilient to the scenarios used to model extreme yet plausible market developments. 
However, the report also highlighted that a significant part of the protection CCPs are equipped with is given by the resources provided by non-defaulting CMs, 
which are in turn at risk of facing significant losses. In severe scenarios, this could trigger second round effects via additional losses at CCP level and the default of additional CMs. 
This is why ESMA recommended CCPs to carefully evaluate the creditworthiness of CMs, as well as their potential exposures due to their participation to other CCPs. 

In this work we address this request by developing a network-based framework for CCPs stress tests that allows to assess contagion effects triggered by different initial shocks 
that propagate through credit and liquidity contagion channels---according to the dynamics proposed by \citet{Lau2009} and later developed by \citet{Cimini2016}.
The model we propose aims at overcoming the current definition of CCPs stress tests used to determine the size of their default funds, 
by considering spillover and contagion effects amongst CMs that indeed should be taken into account when putting aside default resources.
We quantitatively challenge the {\em cover 2} rule to assess the systemic losses it may generate, and whether they are comparable to a distributed shock. 
Specifically, instead of fixing {\em ex-ante} a number of CMs that might default at the same time, we follow an {\em ex-post} approach and determine how many CMs 
would be affected by an initial stress hitting the system and reverberating within CMs. Indeed, the propagation of the initial distress can lead to total losses 
that might be larger than those estimated by the {\em cover 2} requirement. Our stress test methodology is based on the assessment of the financial positions and of the interconnections between CMs 
participating in the market cleared by a given CCP. In particular, we propose a network-based approach for modeling the links between CMs, in order to assess the resilience of the considered market 
to a number of possible financial shocks (namely, idiosyncratic, macroeconomic, price, credit and liquidity shocks).
Given the difficulties, at CCP level, to collect the data needed to assess the exposures of its CMs to other CCPs, here we focus on a single cleared market. 
Thus, without loss of generality, we apply the proposed approach to CC\&G, 
the only clearing house authorized in Italy that operates on several markets and asset classes 
(\eg, Fixed Income, Equities and Equity Derivatives) \footnote{For a complete description of CC\&G's current stress test methodology, please refer to 
\url{http://www.lseg.com/areas-expertise/post-trade-services/ccp-services/ccg/risk-management?preferred-lang=set}.}. 
Note that in order to avoid spill-over effects between CMs trading in different asset classes, CC\&G has a separate default fund for each of these classes. 
Thus we can consider asset classes separately, and in what follows we focus on the Fixed Income class \footnote{The cleared products we consider include 
Italian Government Bonds, Repos and Corporate Bonds traded on the {\em Borsa Italiana} platform.}, which is the most significant in terms of cleared volumes and systemic importance in the Italian financial system. 
Importantly, CC\&G's Fixed Income default fund is generally gauged on a {\em cover 4} basis, \ie, in order to cover the 4 members to which the largest exposure is recorded, 
and is thus more conservative than what prescribed by the {\em cover 2} requirement.

In a nutshell, we find that network effects are relevant and lead to losses comparable to, or even bigger than those resulting from an initial macroeconomic and idiosyncratic shock. 
In general, CC\&G's Fixed Income default fund turns out to be wide enough to cover the uncovered exposures of all defaulting CMs. 
In the scenario corresponding to the default of the two most exposed CMs, initial shocks trigger additional default events, challenging the effectiveness of the {\em cover 2} requirement. 
We support our findings through specific examples, as well as with an exhaustive analysis in the space of the economic parameters of the model.
The paper is organized as follows. In Section \ref{model} we give a step-by-step description of the methodology used in our stress test framework. 
In Section \ref{default} we describe how to assess the {\em cover 2} adequacy under the hypothesis of extreme stress conditions. 
In Section \ref{results} we present and discuss the results of the stress test simulation, and in Section \ref{conclusion} we conclude and outline future perspectives.

\section{Methods}\label{model}

As stated in the Introduction, without loss of generality we consider a single market cleared by a single CCP: CC\&G's Fixed Income asset class. 
The market is composed of $N$ CMs, that are mainly banks but can be financial institutions of different kinds. 
For a generic CM $i$ belonging to this set, its financial position at each date $t$ is summarized by the balance sheet identity: 
\begin{equation}\label{eq:balancesheet}
E_i(t)=A_i(t)-L_i(t)=[A_i^{\mbox{\tiny{INT}}}(t)+A_i^{\mbox{\tiny{OTH}}}(t)]-[L_i^{\mbox{\tiny{INT}}}(t)+L_i^{\mbox{\tiny{OTH}}}(t)],
\end{equation}
where $A_i(t)$ and $L_i(t)$ represent, respectively, total assets and liabilities of the CM. 
We then split $A_i(t)$ into inter-CMs assets $A_i^{\mbox{\tiny{INT}}}(t)$, given by bilateral credits to other CMs, and other assets $A_i^{\mbox{\tiny{OTH}}}(t)$, 
given both by assets to other CMs collateralized by CC\&G and other assets to the rest of the financial system.
Analogously, we separate $L_i(t)$ into bilateral debits from other CMs, $L_i^{\mbox{\tiny{INT}}}(t)$, and other liabilities, $L_i^{\mbox{\tiny{OTH}}}(t)$. 
As clarified in Section \ref{reverberation}, inter-CMs bilateral contracts allow for the propagation of financial distress within CMs. 
The balance sheet identity of eq. (\ref{eq:balancesheet}) defines the equity $E_i(t)$ of CM $i$ as the difference between total assets and liabilities, and $i$ is considered solvent as long as its equity is positive. 
Here, following the literature on financial contagion~\cite{Nier2008,Gai2010,Upper2011}, we take the insolvency condition $E_i(t)\le0$ as a proxy for default. 

\medskip

Given these basic definitions, the operative steps of our stress-test framework are the following: 
\begin{enumerate}[label=\Alph*)]
 \item Use a Merton-like model to obtain daily balance sheet information of CMs from periodic data reports;
 \item Reconstruct the network of inter-CMs bilateral exposures;
 \item Apply a set of initial shocks to the market (idiosyncratic, macroeconomic and on margins posted by CC\&G);
 \item Reverberate the initial shock on the network via credit and liquidity channels, and quantify the final equity loss.
\end{enumerate}

\subsection{Merton-like model for daily balance sheets}\label{merton}

In order to compute the financial position of each CM at each date $t$, given by eq. (\ref{eq:balancesheet}), we have to obtain daily ``dynamic'' values for both total assets and liabilities 
starting from the information disclosed periodically. 
To this end, we use a Merton-like model that estimates the value of a firm's equity according to Black and Scholes option pricing theory. 
The main insight of this approach is that the equity of a firm can be modeled as the price of a call option on the assets of the firm, with a strike price equal to the notional amount of debt 
issued by the company \citep{Merton1974}. Indeed, shareholders are the residual owners of a company: the value of the assets above the debt will be paid out to them, otherwise they will get nothing. 

Here we resort to a variation of the original Merton model where we remove the assumption that default (or insolvency) can only occur at the maturity of the debt. 
We suppose instead that default occurs the first time the firm's total assets fall below the default point, \ie, the notional value of debt.
As suggested by \citet{Bharath2008}, we approximate the face value of the firm's debt with the book value of the firm's total liabilities. 
Pricing techniques for barrier options, whose payoff depends on whether the underlying asset price reaches a certain level during a specified time interval, can be used for our purpose. 
In particular, we consider down-and-out call options, \ie, knock-out call options that cease to exist if the asset price decreases falls below the barrier \citep{Tudela2005}. 
In our framework, the barrier is set equal to the firm's total liabilities and the maturity can be set to $T=1$ year, following \citet{Tudela2005}.

Our approach is based on the assumption that the total assets of a generic firm $i$ follow a geometric Brownian motion
\begin{equation}\label{eq_brown}
dA_i(t) = \mu_i^{(A)} A_i(t) dt + \sigma_i^{(A)} A_i(t) dW, 
\end{equation}
where $\mu_i^{(A)}$ is the expected continuously compounded return on $A_i$, $\sigma_i^{(A)}$ is the volatility of $A_i$ and $dW$ is a standard Wiener process. 
According to the Black and Scholes pricing model, the price of the considered down-and-out call option is given by:
\begin{equation}\label{eq_mert_1}
E_i(t) = N[d_+]A_i(t) - N[d_-]L_i e^{-rT}-N[y]A_i(t) \Biggl(\frac{L_i}{A_i(t)}\Biggr)^{2\lambda} + N[\tilde{y}]L_i e^{-rT}\Biggl(\frac{L_i}{A_i(t)}\Biggr)^{2\lambda-2} ,
\end{equation}
where
\begin{equation*}
d_\pm[A_i(t),\sigma_i^{(A)}] = \frac{1}{\sigma_i^{(A)} \sqrt{T}}\Biggl[\ln\Bigl(\frac{A_i(t)}{L_i}\Bigr)+\Bigl(r\pm\frac{1}{2}(\sigma_i^{(A)})^2\Bigr)T\Biggr],
\qquad \lambda[\sigma_i^{(A)}]=\frac{r}{(\sigma_i^{(A)})^2}+\frac{1}{2},
\end{equation*}
\begin{equation*}
y[A_i(t),\sigma_i^{(A)}]=\frac{1}{\sigma_i^{(A)} \sqrt{T}}\ln\Bigl(\frac{L_i}{A_i(t)}\Bigr)+\lambda\sigma_i^{(A)}\sqrt{T},\qquad
\tilde{y}[A_i(t),\sigma_i^{(A)}]=y-\sigma_i^{(A)}\sqrt{T},
\end{equation*}
$N$ indicates the cumulative function of the standard normal distribution and $r$ is the risk-free rate. 
Moreover, it can be shown that the following relation holds between the equity volatility $\sigma_i^{(E)}$ and the assets volatility $\sigma_i^{(A)}$ \citep{Jones1984}:
\begin{equation}\label{eq_mert_2}
\begin{split}
E_i(t)\sigma_i^{(E)} &= A_i(t)\sigma_i^{(A)}\frac{\partial E_i(t)}{\partial A_i(t)} = \\
&= N[d_+]A_i(t)\sigma_i^{(A)} + N[y]\Biggl[ (2\lambda-1)A_i(t)\sigma_i^{(A)}\Biggl(\frac{L_i}{A_i(t)}\Biggr)^{2\lambda}\Biggr] 
+ N[\tilde{y}]\Biggl[ (2-2\lambda)A_i(t)\sigma_i^{(A)}  e^{-rT}\Biggl(\frac{L_i}{A_i(t)}\Biggr)^{2\lambda-1}\Biggr]\ .
\end{split}
\end{equation}
In the model we adopt, the option value $E_i(t)$ is observed on the market as the total current value of the firm's equity, while its volatility $\sigma_i^{(E)}$ can be easily estimated. On the other hand, 
the unknown variables are the current value of assets $A_i(t)$ and its volatility $\sigma_i^{(A)}$. They can be estimated by inverting the two nonlinear equations (\ref{eq_mert_1}) and (\ref{eq_mert_2}). 

In the implementation of the Merton model, the following input parameters have been used: $E_i(t)$ has been approximated as the firm's market capitalization (if the firm is a listed company) 
or equity (as reported in the last available balance sheet); $\sigma_i^{(E)}$ has been proxied as the volatility of the market capitalization (if the company is listed), otherwise as the volatility 
of a reference index \footnote{If the CM is Italian we use the FTSE Italia All Share Banks Index, otherwise we take the EURO STOXX Banks Eur Index.}, as suggested by \citet{Bharath2008}; 
$L_i$ has been represented as the total liabilities of the firm as reported in the last available balance sheet; $\mu_i^{(A)}$ has been estimated as the annual return on assets of the company.

The model produces the following outputs: $A_i(t)$, an estimate of the firm's assets at time $t$; $\sigma_i^{(A)}$, the volatility of $A_i(t)$; 
$L_i(t)$, an estimate of the firm's liabilities at time $t$; $p_i(t)$, the default probability of the firm. 

Once we have obtained daily balance sheet entries for each CM, we can also determine daily values for their inter-CMs assets and liabilities. 
Given that most of the CMs are banks, we can proxy $A_i^{\mbox{\tiny{INT}}}(t)$ and $L_i^{\mbox{\tiny{INT}}}(t)$ with interbank assets and liabilities as reported on the balance sheet. 
Then we use the approach of \citet{Nier2008} where, for each CM, the proportion of interbank assets (liabilities) over total assets (liabilities) remains constant over time
\footnote{For those CMs that do not disclose interbank figures in their balance sheets, 
we determine daily values of interbank assets and liabilities by using an inverse logarithmic transformation: 
$$\log A_i(t) = \alpha_A + \beta_A \log A_i^{\mbox{\tiny{INT}}}(t) \qquad\qquad \log L_i^{tot}(t) = \alpha_L + \beta_L \log L_i^{\mbox{\tiny{INT}}}(t).$$
Parameter values for the above equation are obtained by fitting quarterly balance sheet data of a pool of banking sector firms, for which both total and interbank assets/liabilities are available. 
Fit results in our case read: $\alpha_A=1.7(0)$, $\beta_A=0.8(1)$, $R^2=0.84$; $\alpha_L=1.0(6)$, $\beta_L=0.9(3)$, $R^2=0.90$.}. %(see Fig. \ref{figS1}).}.

\subsection{The network of inter-CMs exposures}\label{interbank_network}

We now want to build the market of inter-CMs bilateral exposures. Because of its structure, this market can be properly represented as a directed weighted network, 
whose nodes are the CMs and whose links correspond to the direct credits and debits between the CMs \citep{Iori2008,Barucca2016} which will constitute the ground for the propagation of financial distress. 
Indeed, for each CM $i$ at date $t$, its bilateral inter-CMs assets and liabilities are the aggregates of the individual loans to and borrowings from other CMs. Thus 
\begin{equation}\label{eq.net_ibk}
A_i^{\mbox{\tiny{INT}}}(t)=\sum_j a_{ij}^{\mbox{\tiny{INT}}}(t)\qquad\qquad L_i^{\mbox{\tiny{INT}}}(t)=\sum_j l_{ij}^{\mbox{\tiny{INT}}}(t), 
\end{equation}
where $a_{ij}^{\mbox{\tiny{INT}}}(t)$ is the amount of the loan granted by $i$ to $j$ at $t$, which represents an asset for $i$ and a liability for $j$: 
$l_{ji}^{\mbox{\tiny{INT}}}(t)\equiv a_{ij}^{\mbox{\tiny{INT}}}(t)$ $\forall i,j$. 
These amounts represent the weighted links of the interbank network, for which however we have no information. 
To overcome this limitation, we resort to the two-step inference procedure introduced by \citet{Cimini2015a} \footnote{Several methods exist to infer the structure of a network of bilateral exposures 
\citep{Wells2004,Upper2011,Mastromatteo2012,Baral2012,Drehmann2013,Halaj2013,Anand2014,Montagna2014,Peltonen2015}. The advantage of the method we use here \citep{Cimini2015a,Cimini2015b} is that of deriving 
connection probabilities from straightforward principles and of obtaining a network with the correct density of links.} \footnote{Note that adding the layer of CMs' shareholding of other publicly traded CMs 
to the network is in principle possible, but has a negligible impact on our results as the overall volume of shareholding participation within Italian banks is much lower than the total inter-CMs assets (or liabilities) volume.} 
to reconstruct the network. 
We have:
\begin{equation}\label{eq:weight}
a_{ij}^{\mbox{\tiny{INT}}}(t)=\frac{z^{-1}+A_i^{\mbox{\tiny{INT}}}(t)\,L_j^{\mbox{\tiny{INT}}}(t)}{C^{\mbox{\tiny{INT}}}(t)}\,\omega_{ij}(t),\qquad \omega_{ij}(t)=
\begin{cases}
1\quad\mbox{with probability }p_{ij}(t)=\{[z\,A_i^{\mbox{\tiny{INT}}}(t)\,L_j^{\mbox{\tiny{INT}}}(t)]^{-1}+1\}^{-1}\\
0\quad\mbox{otherwise}
\end{cases}
\end{equation}
where $\omega_{ij}(t)$ denotes the presence of the link, $C^{\mbox{\tiny{INT}}}(t)=\sum_iA_i^{\mbox{\tiny{INT}}}(t)$ is the total volume of the inter-CMs market, 
and $z$ is a parameter that controls for the density of the network. Here we set $z$ to have a network density of 5\% like the one observed in the Italian interbank market e-MID 
at a daily aggregation scale \citep{Finger2013}.
Note that contracts can be established not only amongst CMs but also with external firms, and in principle should be included in this network as links pointing out of the system. 
Here, since we proxied inter-CMs trades with interbank trades, and most of CC\&G CMs are Italian banks, 
we use BIS Locational Banking Statistics \footnote{\url{http://www.bis.org/statistics/bankstats.htm}} to confirm that the foreign positions of Italian banks are rather small 
(amounting to roughly 10\% of the total exposure) and can thus be neglected in our analysis.

\subsection{Initial Shocks}\label{initial_shocks}

We now model the initial shock to be applied to the system. For each CM $i$ at date $t$, we decrease its equity by:
\begin{equation}\label{eq:shock_shape}
S_i(t)=\bigl[\phi\xi_i(t)+(1-\phi)\bigr]\chi (t) E_i (t)+\psi_i (t) \max \{ M_i^{\mbox{\tiny{STR}}}(t)-M_i(t),0\}.
\end{equation}
The first term of the right-hand side of eq. (\ref{eq:shock_shape}) is the \textit{exogenous} shock on the assets of CM $i$ which we model, in line with \citet{Manna2012}, with an idiosyncratic and a macroeconomic 
component: given that $\phi\in[0,1]$ and $\xi_i(t)$ is a Poisson random variable with mean one, this shock results from the combination of a stochastic and a deterministic contribution, 
both of which affect $i$ proportionally to its equity $E_i(t)$ (without loss of generality, here we set $\phi=1/2$). 
The magnitude of the shock is set by the parameter $\chi (t) = x [\sum_i A_i(t)]/[\sum_i E_i(t)]$, where $x$ is the average total contribution of the external shock over the total amount of assets in the system 
\footnote{It is $\langle \sum_i\bigl[\phi\xi_i(t)+(1-\phi)\bigr]\chi(t) E_i (t) \rangle = x \sum_i A_i (t)$.}. 
The second term of the right-hand side of eq. (\ref{eq:shock_shape}) is instead due to CC\&G's stress test methodology, in which margins are stressed in order to estimate losses arising 
under ``extreme but plausible" market conditions, represented by historical or simulated yield scenarios that are more conservative than those covered by ordinary margins 
\footnote{In this framework, stressed margins have been calculated under the most conservative amongst CC\&G's stress tests scenarios (for a detailed description of CC\&G's stress testing methodologies, 
please refer to CC\&G's website).}. We explicitly consider this term to pose a further stress on equities, that might indeed decrease due to a liquidity pressure 
resulting from higher margins called by CC\&G in stressed conditions. We quantify this shock as the difference between stressed margins 
$M_i^{\mbox{\tiny{STR}}}(t)$ (the theoretical margin amount that the CM would be required to post under the stress scenario) and ordinary margins $M_i(t)$ (the margin amount actually posted by the CM). 
The magnitude of the shock is set by $\psi_i (t) = [E_i(t)]/[\sum_i E_i(t)]$, which rescales the stress due to the increase of margins to values comparable to CMs equities. 
Overall, note that we only consider stresses that are positive and thus cause equity decreases: for each CM $i$ having $S_i(t)<0$, we impose $S_i(t)=0$.

\subsection{Reverberation of Credit and Liquidity Shocks}\label{reverberation}

The financial distress resulting from the initial equity losses can propagate within the network of inter-CMs exposures, eventually becoming amplified and causing additional losses. 
Two main mechanisms are responsible for such a reverberation: {\em credit} and {\em liquidity} shocks \citep{Cifuentes2005,Nier2008,Lau2009,Krause2012}. Indeed, if a generic CM $i$ defaults, two kinds of events occur: 
 \begin{itemize}
  \item {\em credit shock}: CM $i$ fails to meet its obligations, resulting in effective losses for its creditors;
  \item {\em liquidity shock}: other CMs are unable to replace all the liquidity previously granted by $i$, which in turn triggers a fire sale of assets causing effective losses as illiquid assets trade at a discount. 
 \end{itemize}
If another CM defaults because of these losses, a new wave of credit-funding shocks propagates through the market, eventually resulting in a cascade of failures. 
However, these shocks may propagate even if no default has occurred \citep{Battiston2012}, as equity losses experienced by a CM do imply both a decreasing value of its obligations \citep{Bardoscia2015} 
as well as a decreasing ability to lend money to the market \citep{Cimini2016}, because the CM is now ``closer'' to default. This results in potential equity losses for other CMs. 
In order to quantify these potential losses we use the approach of \citet{Cimini2016}, which builds on \citet{Battiston2012}, to obtain probabilities of CMs default (that is, the equity at risk)
by iteratively spreading the individual CMs distress levels weighted by the potential wealth affected. 

In a nutshell, the method works as follows. Assuming that relative changes of equity translate linearly into relative changes of claim values, the resulting {\em impact} of $i$ on $j$ reads:
\begin{equation}\label{eq:impact}
\lambda\Lambda_{ji}(t)+\rho\gamma(t)\Upsilon_{ji}(t)=
\frac{\lambda a_{ji}^{\mbox{\tiny{INT}}}(t) + \rho\gamma(t) a_{ij}^{\mbox{\tiny{INT}}}(t)} {E_j(t)},
\end{equation}
where $0\le\lambda\le1$ is the parameter setting the amount of loss given default, $0\le\rho\le1$ sets the fraction of lost liquidity that has to be replenished by asset sales 
and $\gamma(t)$ quantifies asset depricing during fire sales. The dynamics of shock propagation then consists of several rounds $\{n\}$, 
and the variables involved are the levels of financial distress of each CM $i$ at each iteration $n$, given by the relative changes of equity
\begin{equation}
h_i^{[n]}(t)=1-E_i^{[n]}(t)/E_i^{[0]}(t).
\end{equation}
By definition, $h_i(t)=0$ when no equity losses occurres for $i$, $h_i(t)=1$ when $i$ defaults, and $0<h_i(t)<1$ in general.
\begin{itemize}
 \item At step $n=0$ there is no distress in the system, hence $E_i^{[0]}(t)\equiv E_i(t)\Rightarrow h_i^{[0]}(t)=0$ $\forall i$; 
 \item At step $n=1$ we apply the initial shock of eq. (\ref{eq:shock_shape}), so that $E_i^{[1]}(t)\equiv E_i(t)-S_i(t)\Rightarrow h_i^{[1]}(t)=S_i(t)/E_i(t)$ $\forall i$;
 \item Subsequent values of $h$ are obtained by spreading this shock on the system and writing up the equation for the evolution of CMs equity \citep{Bardoscia2015}: 
\begin{equation}\label{eq:acca}
h_i^{[n+2]}(t)=\min\left\{1,\; h_i^{[n+1]}(t)+\sum_{j\in\mathcal{A}[n+1]}[\lambda\Lambda_{ij}(t)+\rho\gamma^{[n]}(t)\Upsilon_{ij}(t)]\,[h_j^{[n+1]}(t)-h_j^{[n]}(t)]\,e^{-(n-n_j)/\tau}\right\}. 
\end{equation}
\end{itemize}
In the above expression, $\mathcal{A}[n+1]=\{j:h_j^{[n]}(t)<1\}$ is the set of CMs that have not defaulted up to iteration $n$ and that thus can still spread their financial distress. 
Shocks are exponentially damped with $e^{-(n-n_j)/\tau}$ where $n_j:h_j^{[n_j]}(t)>0, h_j^{[n_j-1]}(t)=0$ is the iteration when $j$ first becomes distressed and $0\le\tau<\infty$ is the damping scale 
setting the mean lifetime of the shocks. For instance, $\tau=0$ means that each CM spreads its distress only the first time it becomes distressed and $\tau\to\infty$ that CMs always propagate received shocks until they default 
\citep{Bardoscia2015}. The fire sale devaluation factor is $\gamma^{[n]}(t)=\{C^{\mbox{\tiny{INT}}}(t)/[\rho Q^{[n]}(t)]-1\}^{-1}$ and the aggregate amount of interbank assets potentially to be liquidated at $n$ is 
$Q^{[n]}(t)=\sum_{j\in\mathcal{A}[n+1]}\sum_k a_{jk}^{\mbox{\tiny{INT}}}(t) [h_j^{[n+1]}(t)-h_j^{[n]}(t)]\,e^{-(n-n_j)/\tau}$ \citep{Cimini2016}. 
The above described dynamics stops at $n^*$ when no more CMs can propagate their distress. 
The set of {\em vulnerabilities} $\{h_i^{[*]}(t)\}_{i\in N}$ then quantifies CMs final potential equity losses.

\section{{\em Cover 2} assessment}\label{default}

We now model a scenario in which, due to ``extreme but plausible'' market conditions, the two CMs to which CC\&G has the largest exposure become insolvent at the same time
\footnote{The two CMs are selected through ordinary CC\&G's stress test procedures.} (corresponding to the {\em cover 2} requirement). 
The rationale behind this analysis derives from the prescription of EMIR to gauge default funds so as to cover at least losses stemming from the default of two CMs, as mentioned in the Introduction. 

In practical terms, this analysis consists of the same steps described in Section \ref{model}, except that the initial equity shock is no longer described by eq. (\ref{eq:shock_shape}). 
Instead, we use an initial condition given by the default of the two selected CMs $\tilde{i}$ and $\tilde{j}$: 
\begin{equation}\label{eq:shock_whatif}
E_{\tilde{i}}^{[1]}(t)=E_{\tilde{j}}^{[1]}(t)=0\Rightarrow h_{\tilde{i}}^{[1]}(t)=h_{\tilde{j}}^{[1]}(t)=1 \qquad E_k^{[1]}(t)=E_k ^{[0]}(t)\Rightarrow  h_k^{[1]}(t)=0 \quad \forall k \neq \tilde{i},\tilde{j}.
\end{equation}
Shocks then propagate as explained in Section \ref{reverberation}, and we obtain the set of vulnerabilities for the {\em cover 2} assessment.

\section{Results and Discussion}\label{results}

We now present the results of our network-based stress test methodology for different choices of model parameters. 
Results are obtained as averages over an ensemble of 1000 realizations of steps B, C, D of our framework.

Input data for the stress-test exercise is CMs' most recent balance sheet information (as available as of September 2016), as well as CC\&G's Fixed Income margins collected from July to October 2016. 
Results presented in this paper refer to a specific date in such a range, yet this particular choice does not alter significantly our results because 
in that period we observe a small relative change of 10\% for the total uncovered exposures of the two most exposed CMs, and of 11\% for the default fund 
\footnote{These variations are in line with the ones obtained considering a longer time period, \ie, from September 2015 to October 2016, for which 
the change of the total uncovered exposures of the two most exposed CMs is 15\%, while that of the default fund is 10\%.}. 
In what follows, we omit the explicit dependence of quantities on the chosen date $t$.

\subsection*{Dynamics of individual CMs vulnerabilities} 

We start with a specific example in order to properly understand the results of our stress-test simulation. 
Hence we set $\lambda=0.6$ (the average loss given default value observed for CC\&G's CMs) and $\rho=0.6$ (set homogeneously with $\lambda$). 
We are interested in comparing, for each CM $i$, three quantities:
\begin{itemize}
 \item $h_i^{[1]}$, the vulnerability (relative equity loss) given by the initial shock;
 \item $h_i^{[2]}$, the vulnerability after the first round of shock propagation on the inter-CMs network 
 (note that stopping the dynamics at early rounds like $n=2$ models corrective actions that are reasonably expected to be implemented by regulators to reduce losses);
 \item $h_i^{[*]}$, the vulnerability after propagation of shocks on the network is exhausted. 
 Here we use the worst-case reverberation $\tau=\infty$, however we observe that the dynamics makes very few iterations to get very close to the stationary configuration 
 (this also means that the model is not highly sensitive to the value of $\tau$ unless it becomes very close to zero).
\end{itemize}
Obviously, it is $h_i^{[1]}\le h_i^{[2]}\le h_i^{[*]}$. 

\begin{figure}[t!] 
\includegraphics[width=0.475\textwidth]{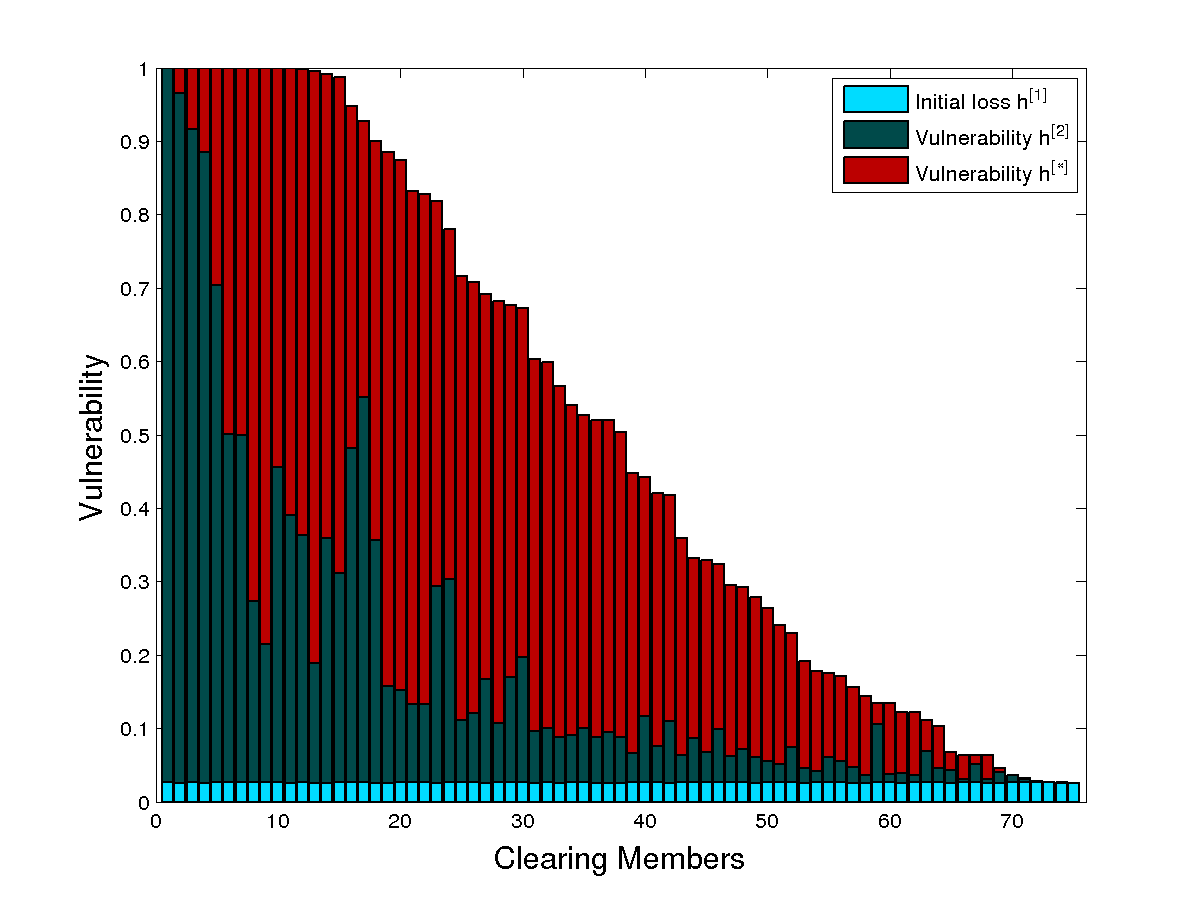}\\
\includegraphics[width=0.475\textwidth]{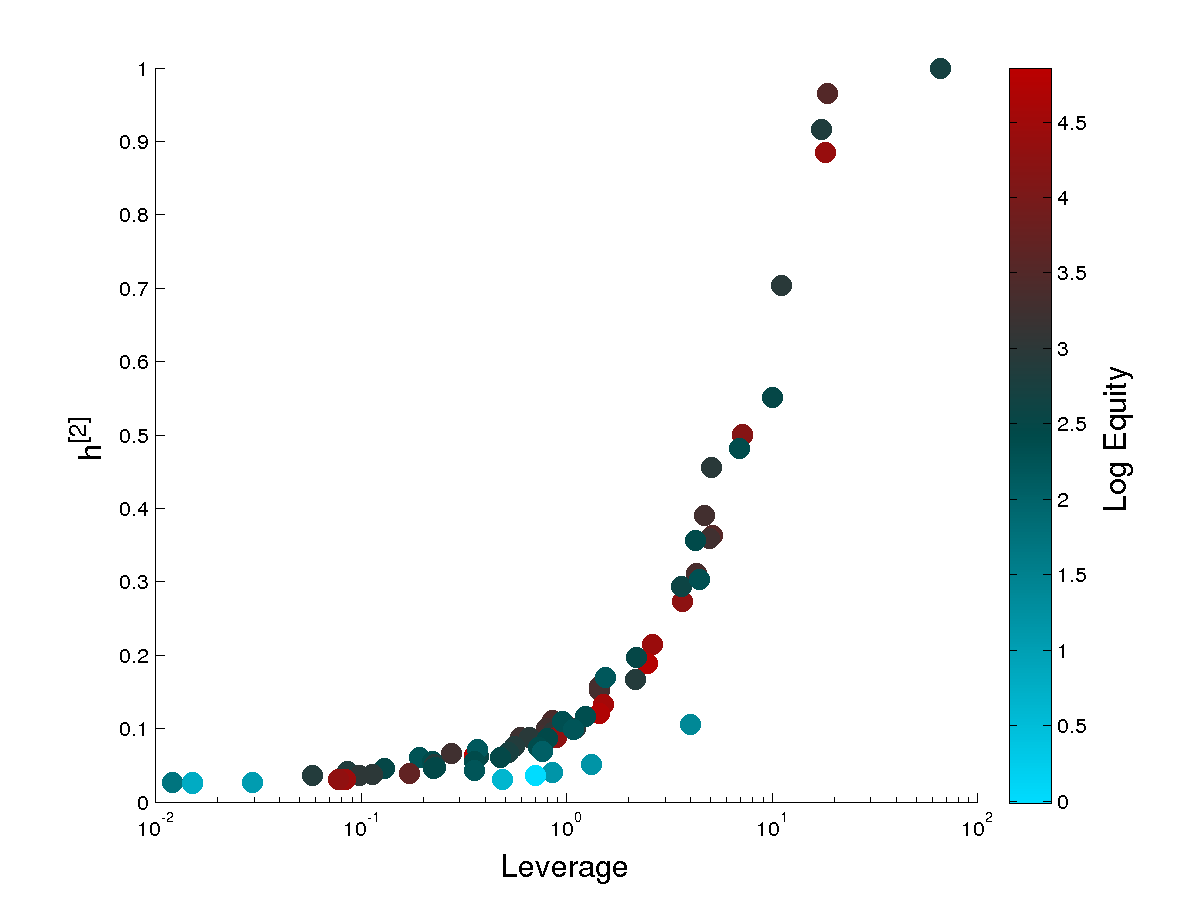}
\includegraphics[width=0.475\textwidth]{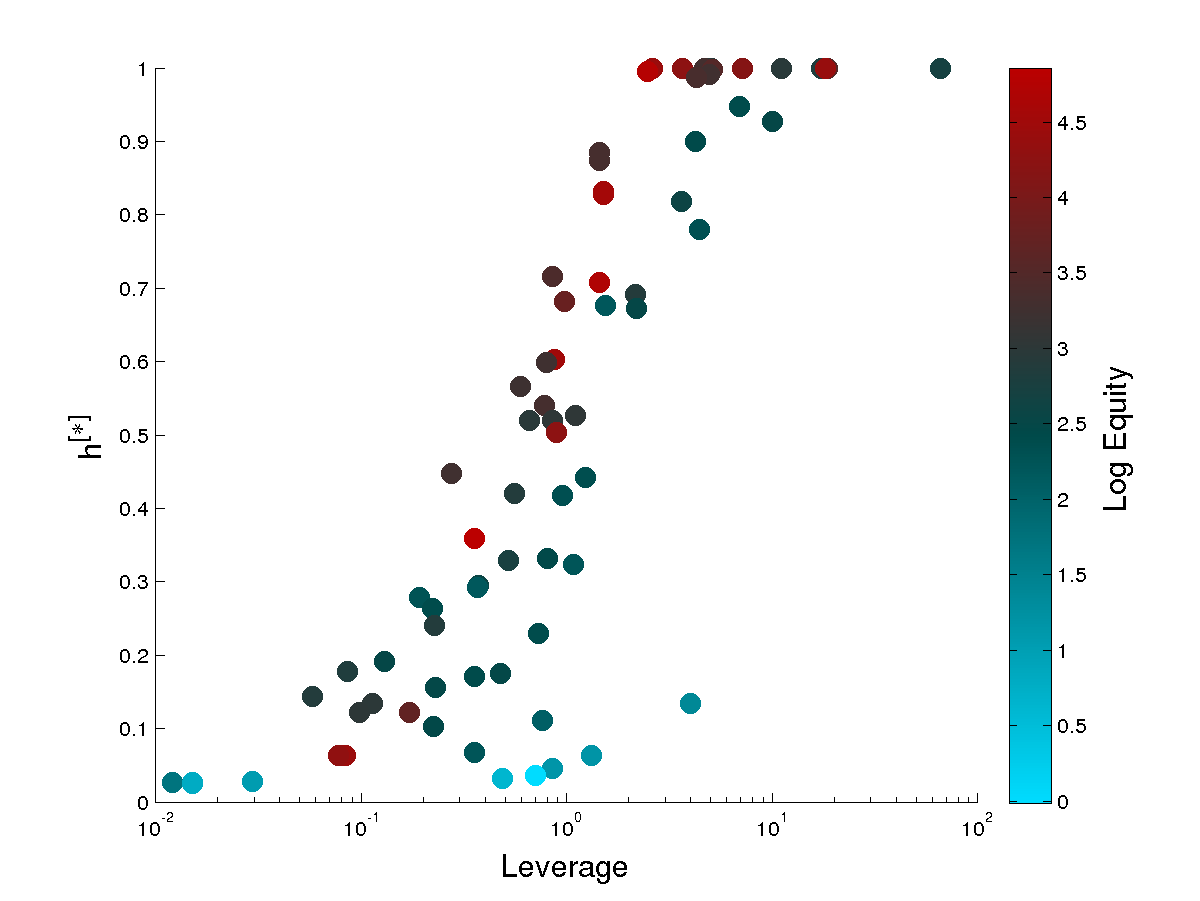}
\caption{{\bf Individual CMs vulnerabilities for distributed initial shocks}, obtained for each CM from eq. (\ref{eq:shock_shape}) for $x=10^{-3}$.
The histogram in the top panel shows for each CM $i$ the triplet $h_i^{[1]},h_i^{[2]},h_i^{[*]}$, ordered by final equity loss. Bottom panels show instead scatter plots of $h_i^{[2]}$ (left) and of $h_i^{[*]}$ (right) 
as a function of the inter-CM leverage $\Lambda_i=A_i^{\mbox{\tiny{INT}}}/E_i$ (each bubble is a CM, colored according to its initial equity).}
\label{fig1}
\end{figure}

Figure \ref{fig1} shows these values for the configuration of distributed initial shocks given by eq. (\ref{eq:shock_shape}) with $x=10^{-3}$, 
which corresponds, on average, to a $2.6\%$ initial equity loss for each CM \footnote{The rationale behind this choice of $x$ is detailed in the next subsection.}. 
We see that initial shocks are very small compared to the magnitude of first-round and total network losses, showing the relevance of network effects. 
Yet, there is a significant heterogeneity in such losses for the various CMs. Overall,
the system is rather stable after the first round of shock propagation ($n=2$), and CMs with greater inter-CM leverage values $\Lambda_i=A_i^{\mbox{\tiny{INT}}}/E_i$ are generally characterized by greater vulnerabilities. 
However, as shocks keep propagating ($n\to n^*$), several CMs get very close to default. The dependence of vulnerability on leverage is still evident, and the critical value $\Lambda=1$ emerges 
as the threshold that separates a regime of low losses from high losses---indeed, a CM $i$ can default in this scenario only if $\Lambda_i\ge1$ \citep{Bardoscia2015,Cimini2016}. 
If we consider defaulted CMs (\ie, those with vulnerability equal to 1) in this extreme scenario, their total uncovered exposure (\ie, the additional guarantees, above what already posted, needed to cover 
a hypothetical stressed margins call) as calculated in CC\&G's stress test is \euro 3.0 billions, which is widely covered by CC\&G's total default fund at the date (\euro 3.5 billions). 
We thus confirm CC\&G's default fund conservativeness since, even under the assumption of an unlimited shocks reverberation, the remaining exposure of all defaulted CMs would be covered.

\begin{figure}[t!] 
\includegraphics[width=0.475\textwidth]{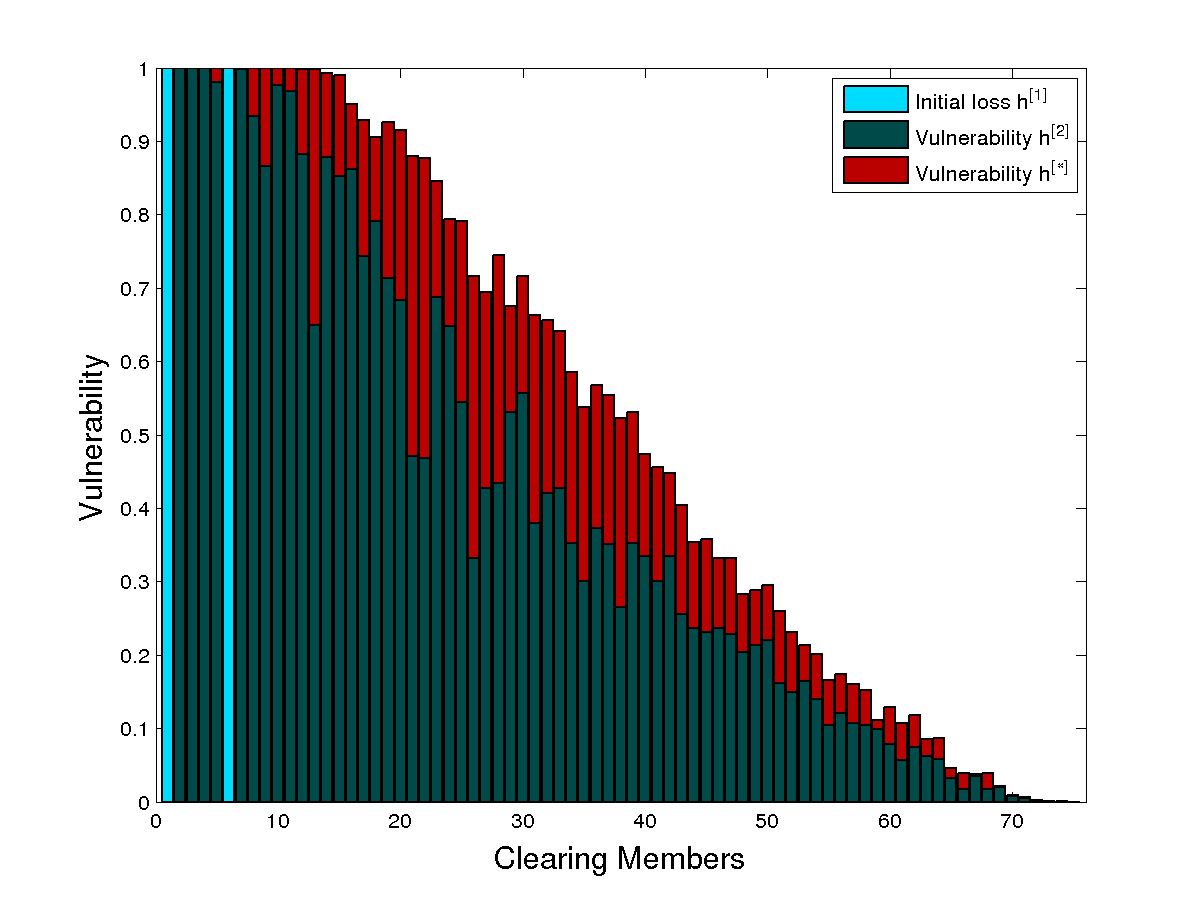}\\
\includegraphics[width=0.475\textwidth]{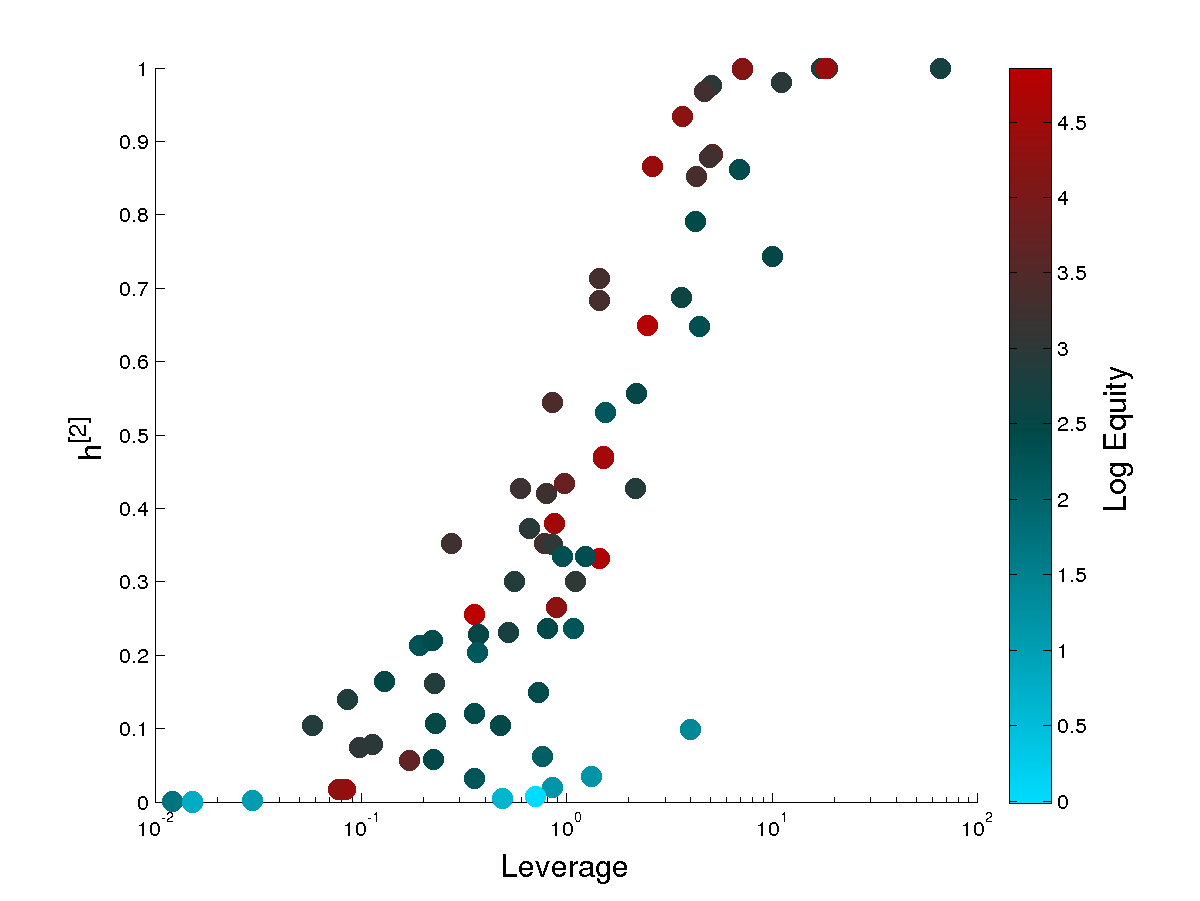}
\includegraphics[width=0.475\textwidth]{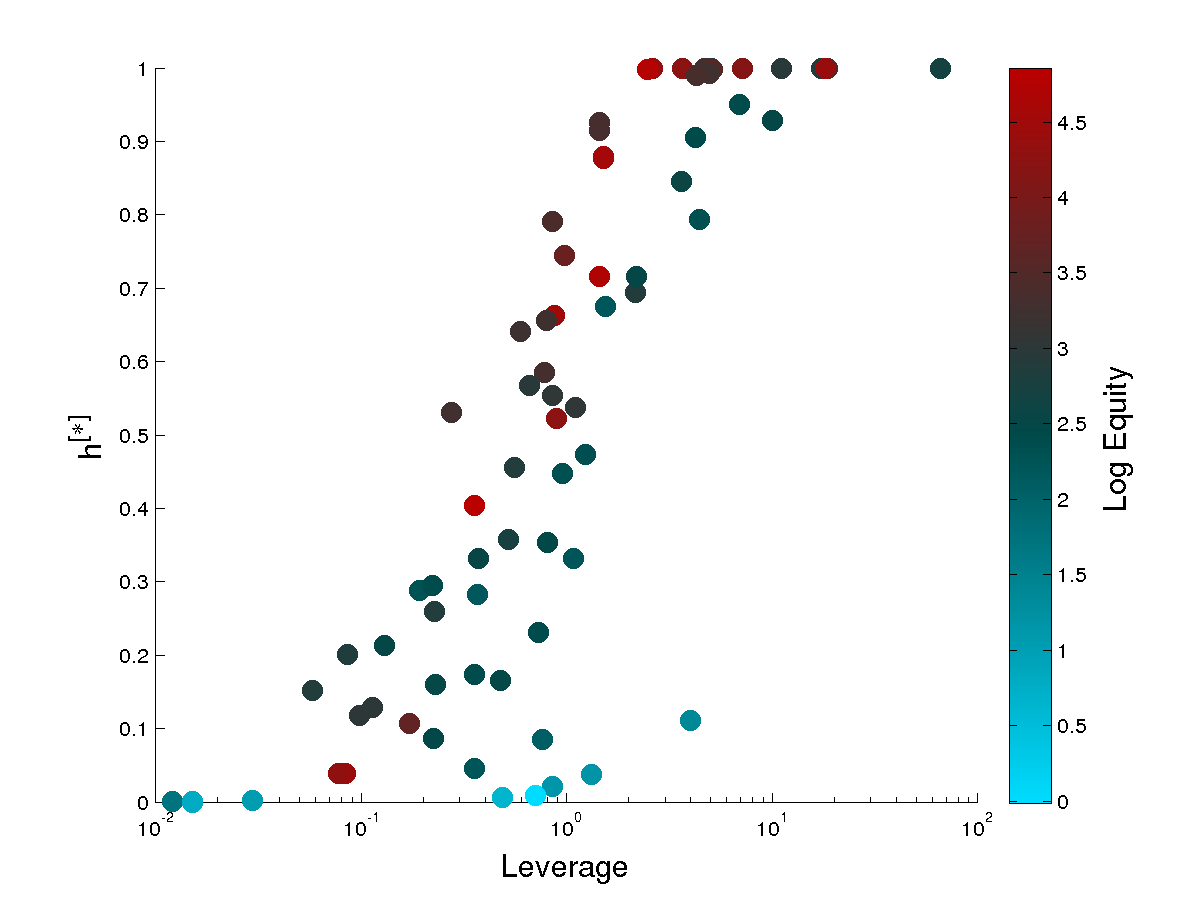}
\caption{{\bf Individual CMs vulnerabilities for {\em cover 2} initial shocks}, obtained from eq. (\ref{eq:shock_whatif}). 
The histogram in the top panel shows, for each CM $i$, the triplet $h_i^{[1]},h_i^{[2]},h_i^{[*]}$, in the same order of the histogram in Fig. \ref{fig1}. 
Bottom panels show instead scatter plots of $h_i^{[2]}$ (left) and of $h_i^{[*]}$ (right) as a function of the inter-CM leverage $A_i^{\mbox{\tiny{INT}}}/E_i$ (each bubble is a CM, colored according to its initial equity).}
\label{fig2}
\end{figure}

Figure \ref{fig2} shows instead results for the configuration of {\em cover 2} initial shocks given by eq. (\ref{eq:shock_whatif}). 
Note that the total initial loss in this case (which stems from the default of the two most exposed CMs) is equivalent to a $3.0\%$ initial equity loss for each CM. 
Hence, the two scenarios reported in Figures \ref{fig1} and \ref{fig2} are comparable in terms of initial shock magnitude. 
However, the very different initial conditions give rise to very different configurations (and magnitudes of losses) at early stages of the shock propagation dynamics ($n=2$). 
Then, as shocks keep propagating ($n\to n^*$), the system falls into a similar stationary configuration---which basically corresponds to the maximum allowed losses for each CM. 
In this scenario, the total uncovered exposure of all defaulted CMs is \euro 3.2 billions, again comparable to the one obtained in the distributed initial shocks case. 
Thus even in this case CC\&G's default fund, which is gauged far more conservatively than what the {\em cover 2} rule prescribes, is capable of covering the total uncovered exposure of the system.
However, the fact that the default of two CMs can lead to additional defaults suggests that gauging the default fund solely on a {\em cover 2} basis might not be conservative enough.

\subsection*{Model dependence on initial shocks and reverberation rounds}

We now wish to study the stability regime of the market with respect to the magnitude of the initial shock and the shock reverberation round. 
To this end, we compute two quantities to measure systemic losses:
\begin{itemize}
 \item {\em residual default fund ratio}: $R_{DF}^{[n]}$, obtained as the fraction of default fund which remains after subtracting the Fixed Income exposures 
of defaulted CMs at reverberation round $n$ (\ie, those with $h^{[n]}=1$);
 \item {\em total relative residual equity}: $R_{RE}^{[n]}=1-\sum_i(E_i^{[1]}-E_i^{[n]})/\sum_i E_i^{[1]}$, namely the ratio of equity which remains after $n$ rounds of shocks reverberation.
\end{itemize}
Note that the first quantity considers overall losses but accounts only for the CMs that actually default, whereas, 
the second quantity takes into account only network losses (by discounting the initial shock) but accounts for each CM irrespectively of its final vulnerability. 
We thus expect a noisier behavior of $R_{DF}$ compared to $R_{RE}$ because of the strict thresholding used in its computation.

\begin{figure}[t!] 
\includegraphics[width=0.475\textwidth]{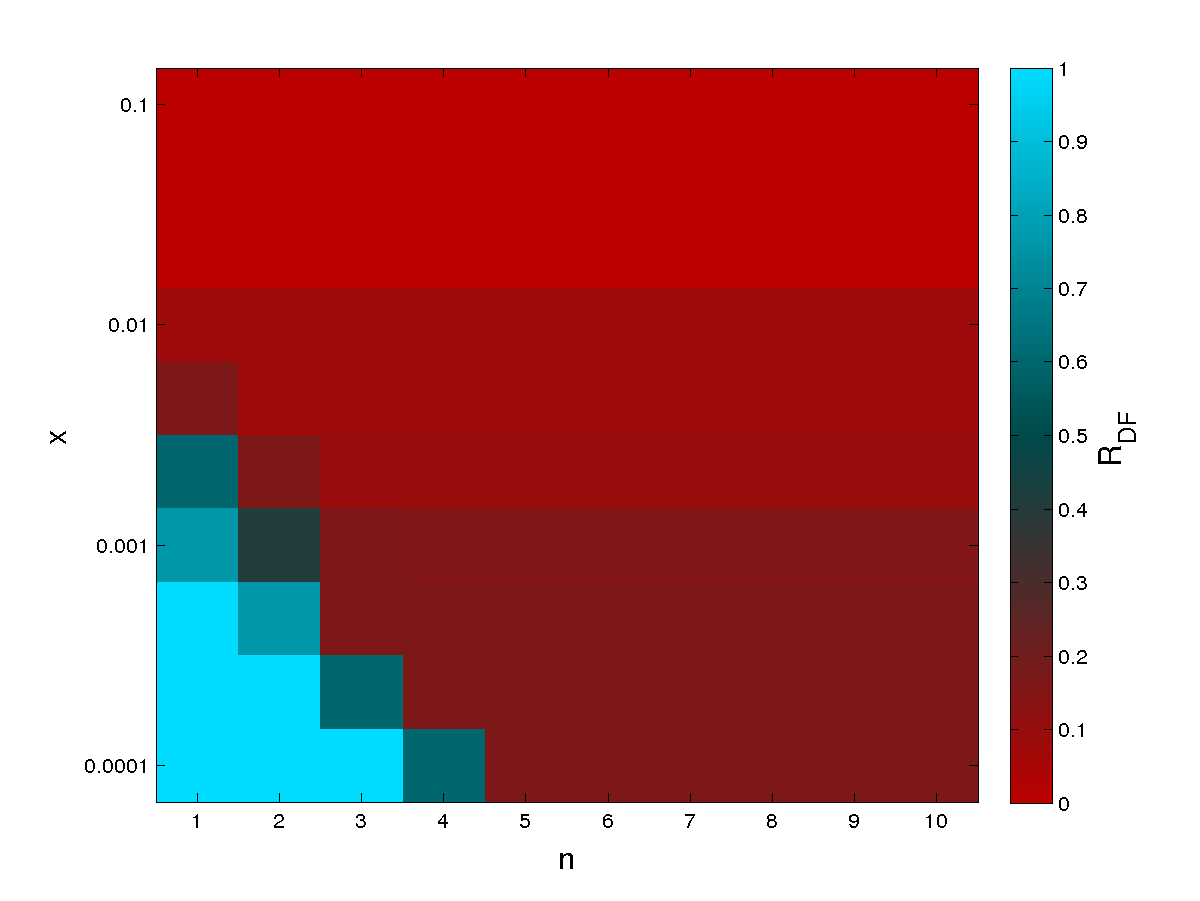}
\includegraphics[width=0.475\textwidth]{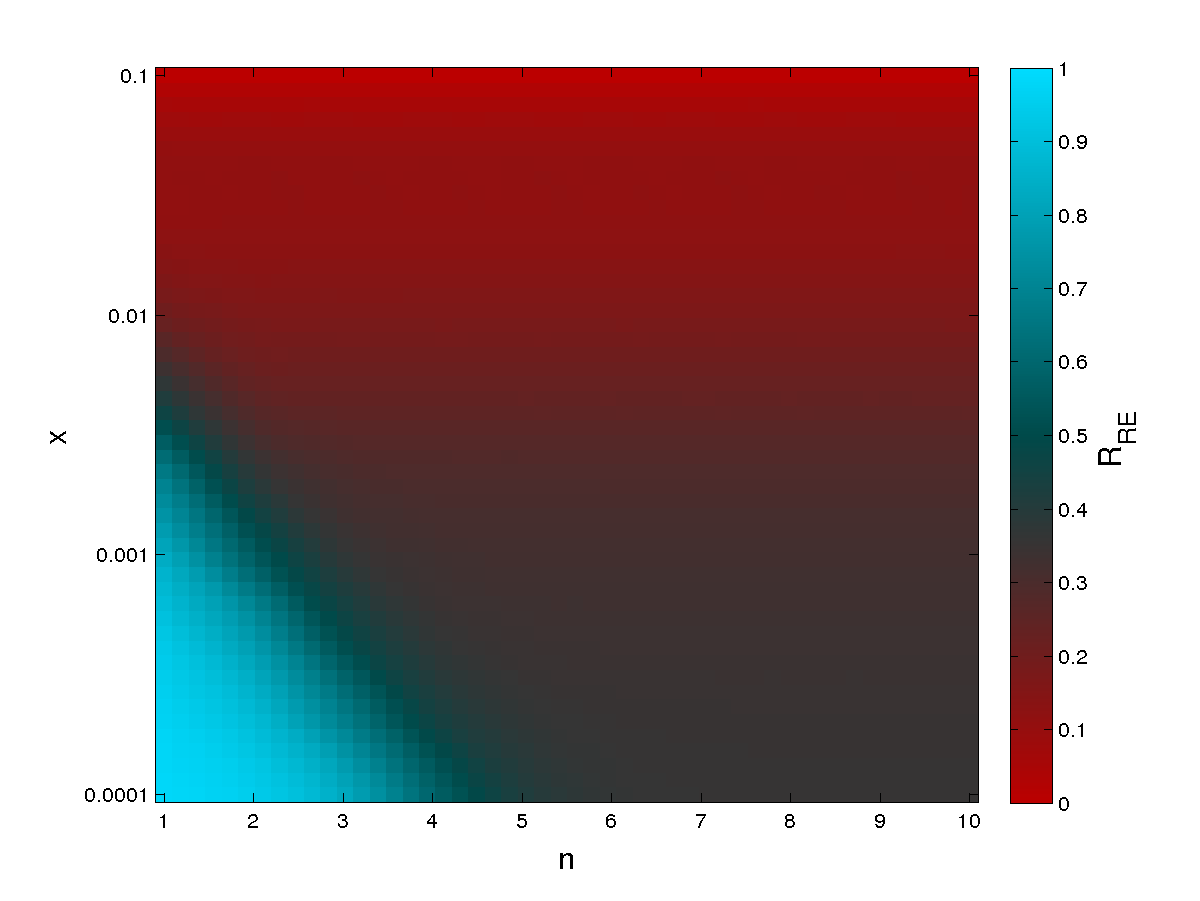}
\caption{{\bf Color map of systemic losses with respect to the macroeconomic initial shocks magnitude $x$ and reverberation round $n$}, measured by $R_{DF}$ (left panel) and $R_{RE}$ (right panel). 
Note that by varying $x$ we consider only the scenario of distributed initial shocks of eq. (\ref{eq:shock_shape}) but, as we have seen in the analysis of Figure \ref{fig2}, 
for large values of $n$, systemic losses are also representative for the {\em cover 2} scenario with similar magnitude of initial shocks. 
Note also that we take $n=10$ as a proxy for $n^*$, as the dynamics of shocks reverberation takes a few iterations for convergence.}
\label{fig3}
\end{figure}

Figure \ref{fig3} shows the color maps of these two quantities as a function of the iteration step of the dynamics ($n$) and of the magnitude of the initial shock ($x$) as per eq. (\ref{eq:shock_shape}). 
We observe a region of small $n$ and $x$ for which systemic losses are rather restrained, and a sharp transition to a region of high losses where the dependence of losses on $x$ dominates. 
Overall, CC\&G's default fund always succeeds in covering all exposures except for unreasonable high values of $x$. 
Indeed, the default fund becomes insufficient ($R_{DF}\to0$) starting from $x\simeq10^{-2}$. 
However, values of $x>10^{-3}$ do not seem conceivable. To have a quantitative estimate for the range of plausible values for $x$, 
we can assume that losses arising from non-performing-loans (NPLs) recorded in the market are likely values of assets losses due to the initial shock. 
In particular, the yearly increase (for 2014/2015) of losses from NPLs for Italian banks goes from 1.37\% (considering also banks reducing these losses) to 2.5\% (considering only losses increases). 
Using these values as initial vulnerabilities and inverting the first term of the right-hand side of eq. (\ref{eq:shock_shape}), we obtain $5 \cdot 10^{-4}<x<10^{-3}$.

\subsection*{Model dependence on credit and liquidity shocks propagation} 

\begin{figure}[t!]
\includegraphics[width=0.475\textwidth]{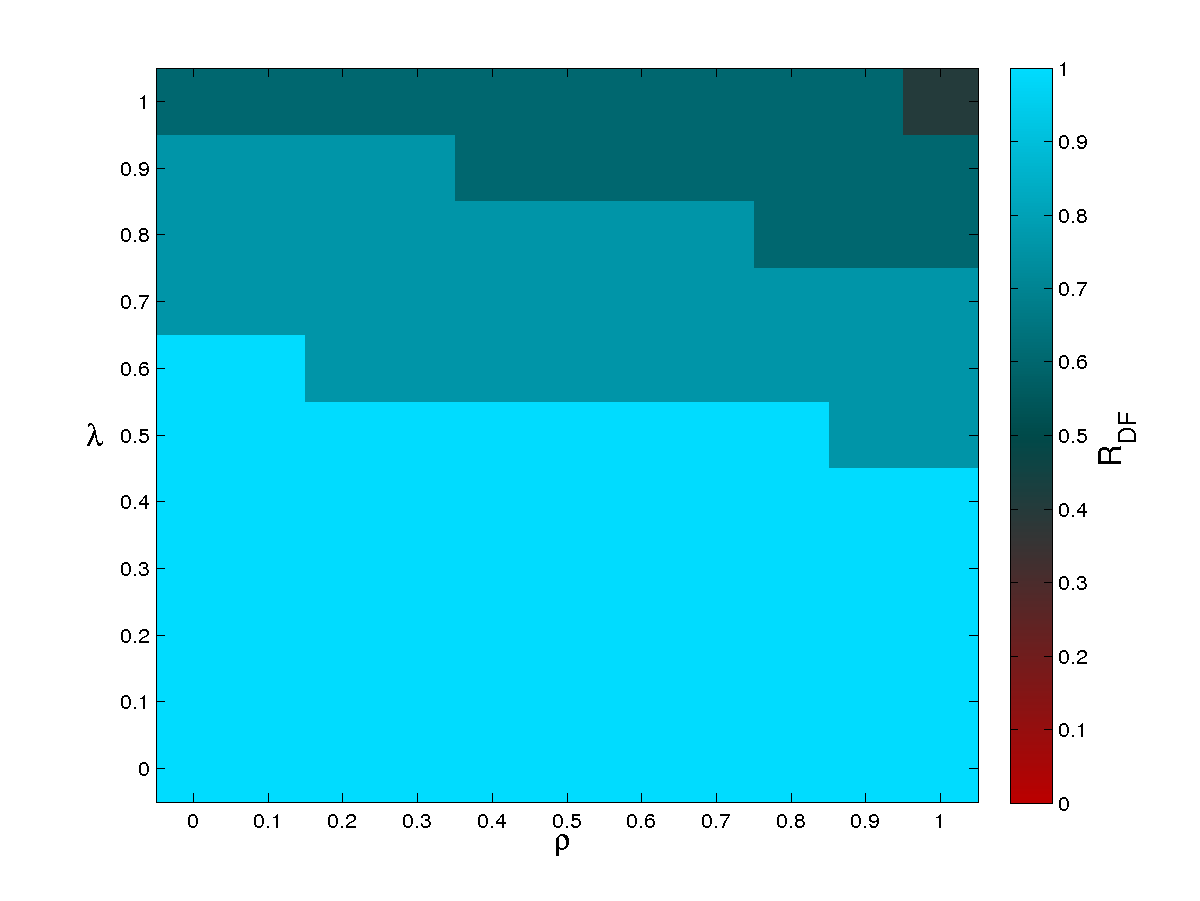}
\includegraphics[width=0.475\textwidth]{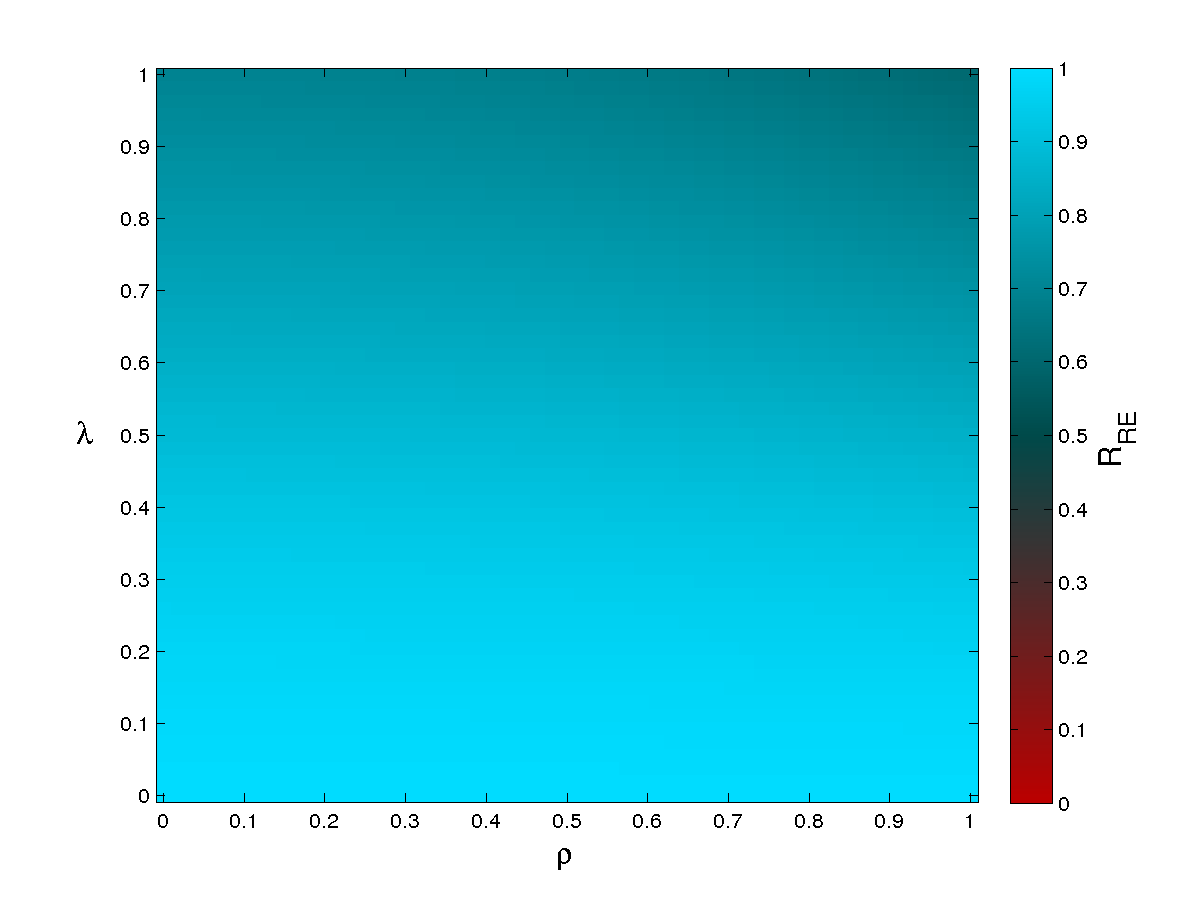}
\includegraphics[width=0.475\textwidth]{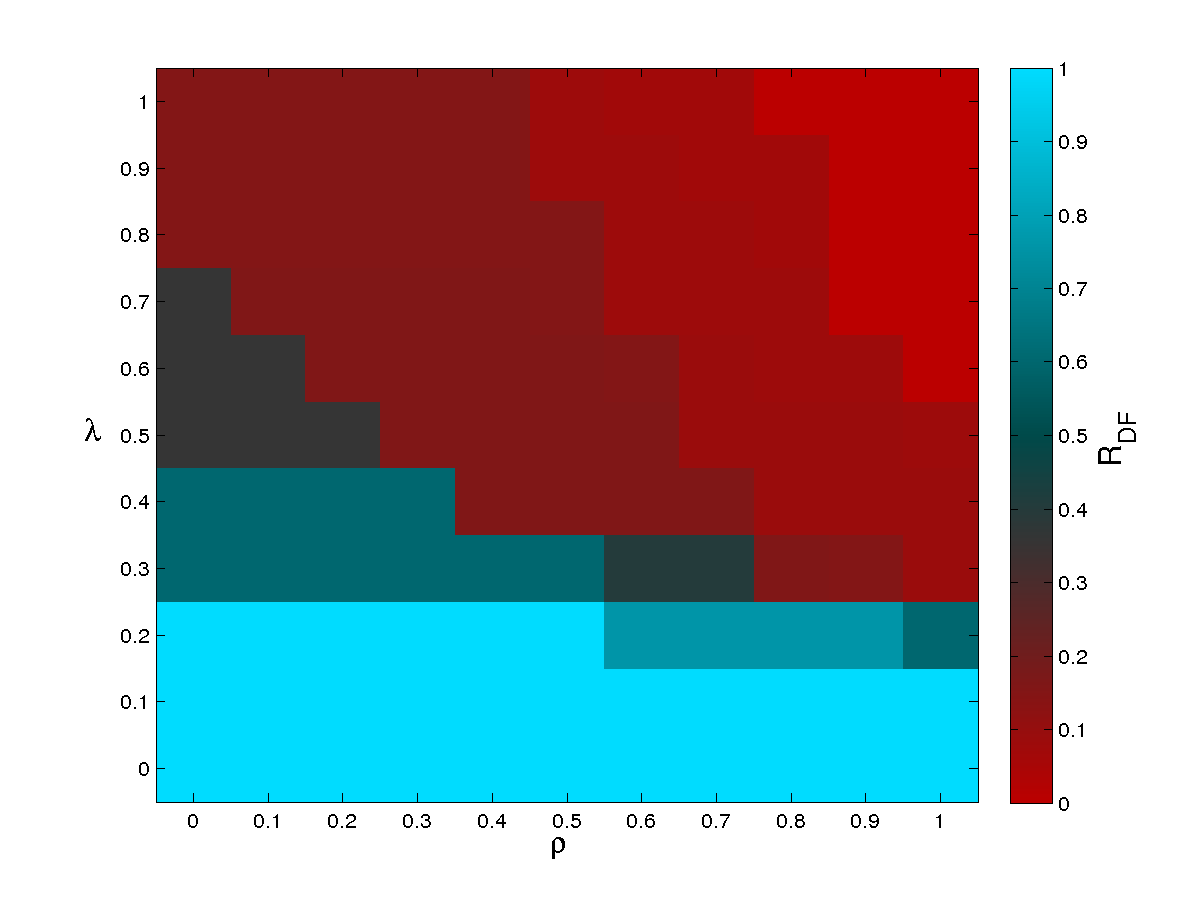}
\includegraphics[width=0.475\textwidth]{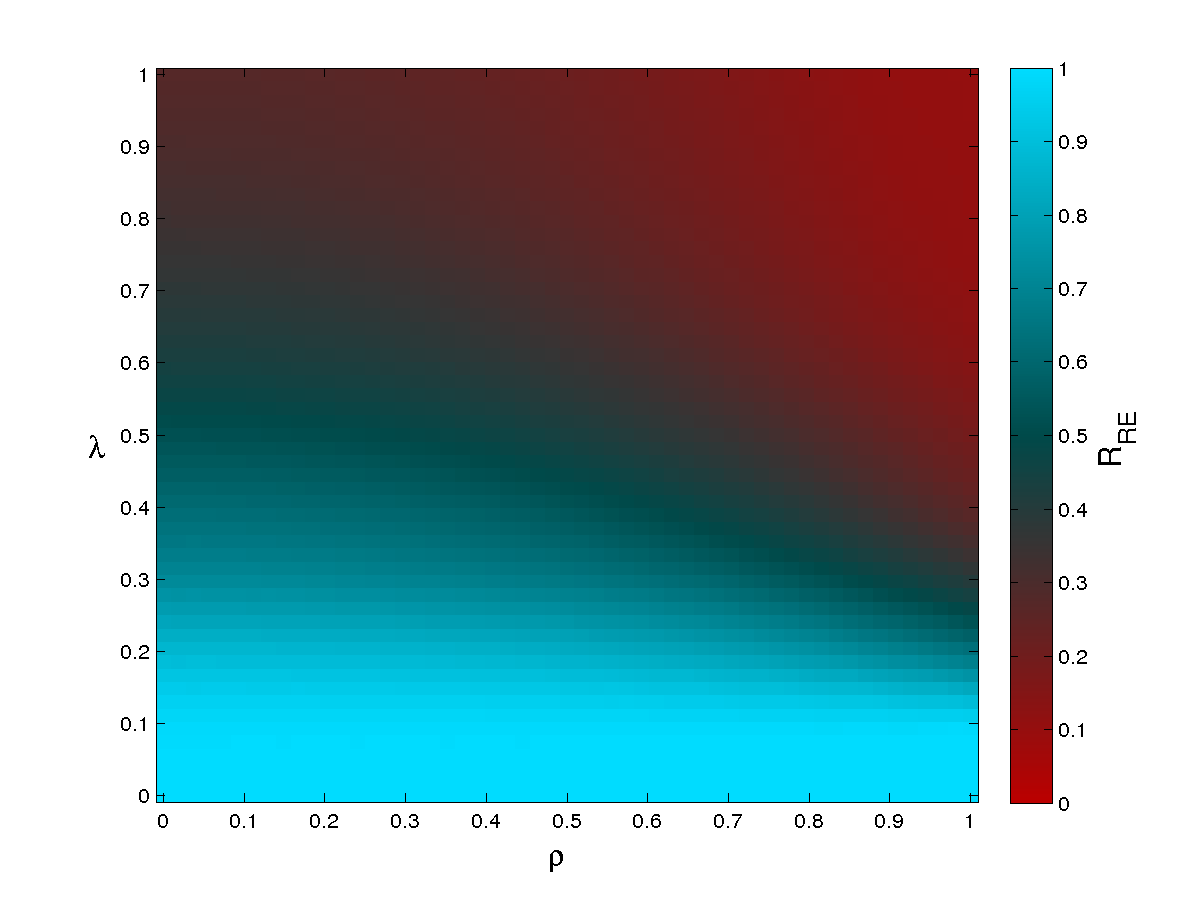}
\caption{{\bf Color map of systemic losses in the distributed shock scenario with respect to the loss given default ($\lambda$) and the lost funding to be replenished ($\rho$)}, 
measured by $R_{DF}$ (left panels) and $R_{RE}$ (right panels), and for $n=2$ (top panels) and $n=n^*$ (bottom panels).}
\label{fig4}
\end{figure}

\begin{figure}[t!]
\includegraphics[width=0.475\textwidth]{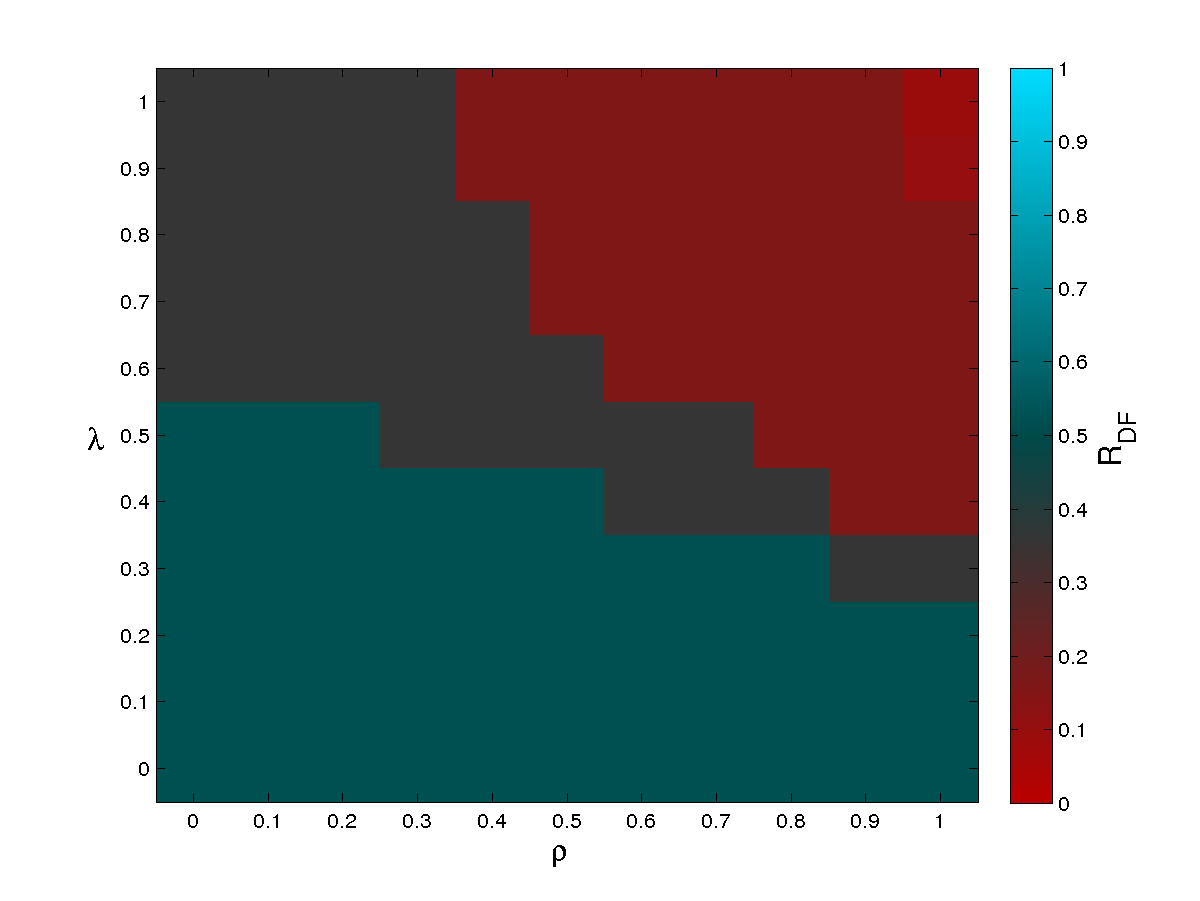}
\includegraphics[width=0.475\textwidth]{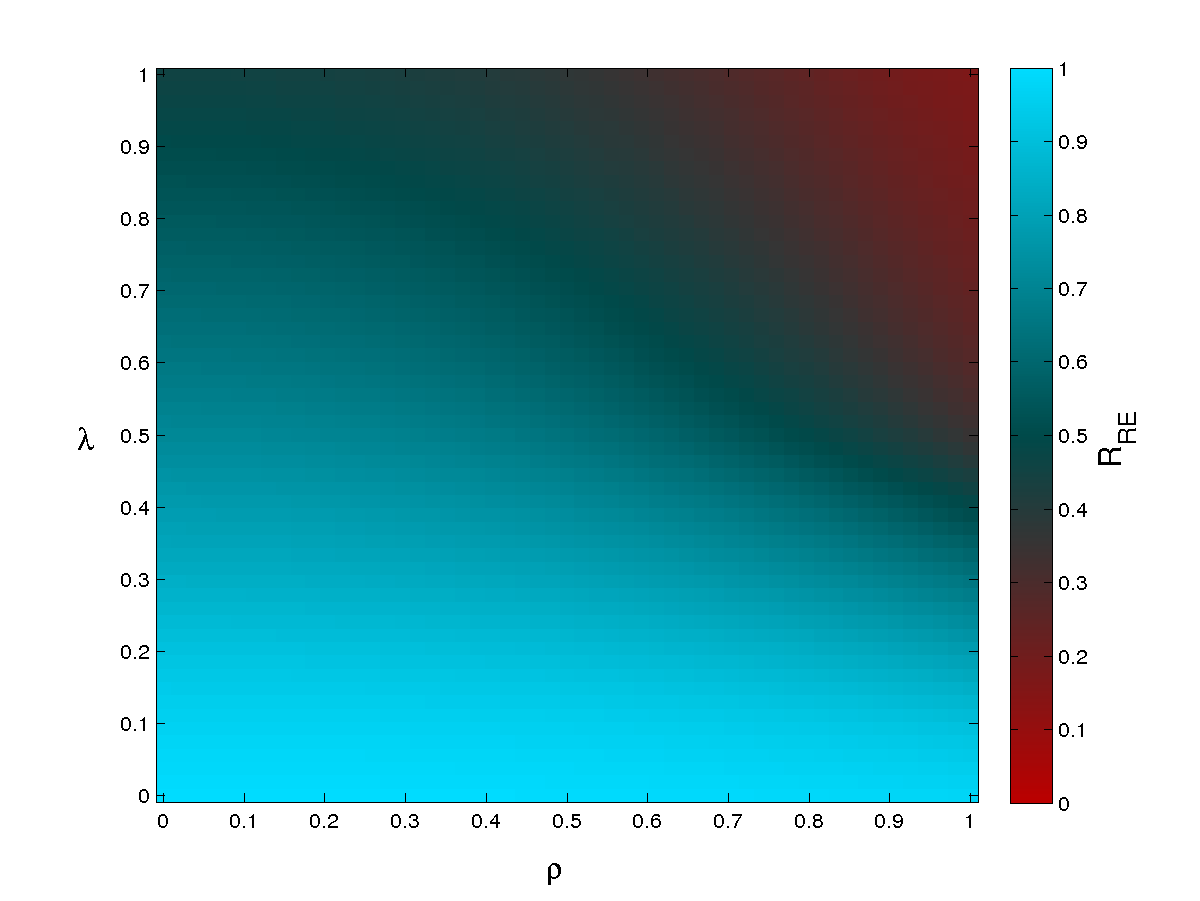}
\includegraphics[width=0.475\textwidth]{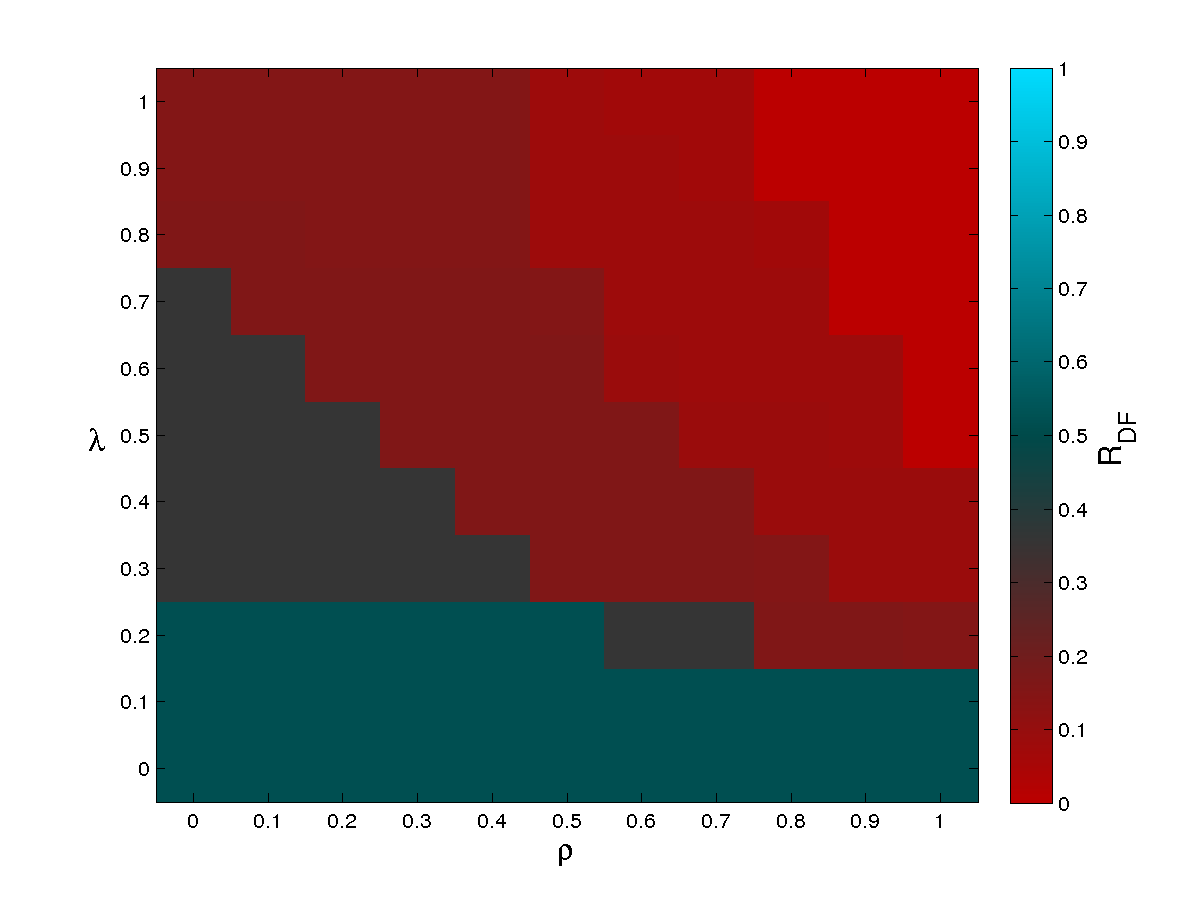}
\includegraphics[width=0.475\textwidth]{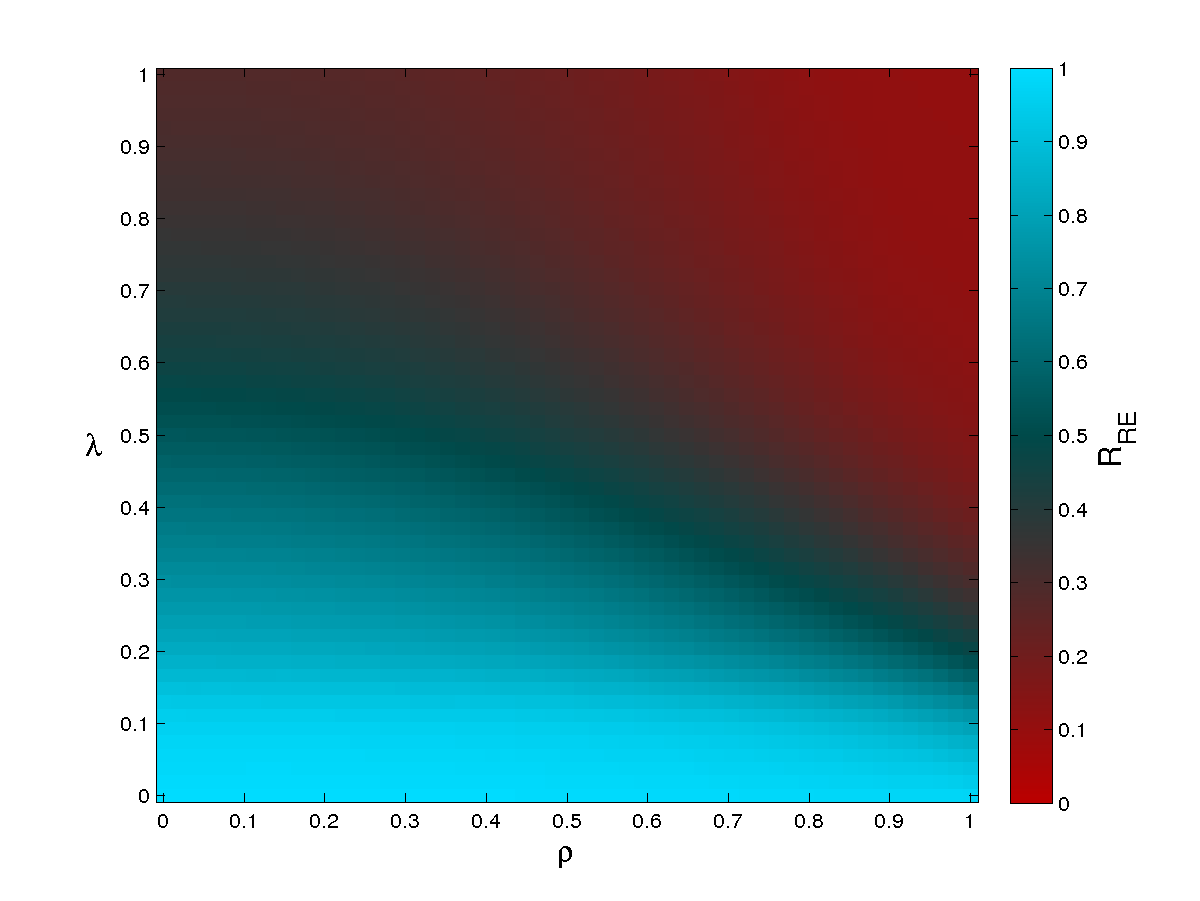}
\caption{{\bf Color map of systemic losses in the {\em cover 2} scenario with respect to the loss given default ($\lambda$) and the lost funding to be replenished ($\rho$)}, 
measured by $R_{DF}$ (left panels) and $R_{RE}$ (right panels), and for $n=2$ (top panels) and $n=n^*$ (bottom panels).}
\label{fig5}
\end{figure}

We finally study the stability regime of the system with respect to the parameters $\lambda$ and $\rho$, setting the intensity of credit and liquidity shocks respectively. 
Here we consider both kinds of initial shocks, \ie, distributed and {\em cover 2} scenarios reported in Figures \ref{fig4} and \ref{fig5} respectively. 

As expected, in general we observe higher losses for greater values of $\lambda$ and $\rho$, and a rather smooth transition between the regimes of low and high losses. 
In the case of distributed shocks (Figure \ref{fig4}, obtained again with $x=10^{-3}$), the first network reverberation of shocks never leads to severe consequences. 
However, if shocks keep propagating the default fund deteriorates significantly, becoming insufficient in the region above the segment delimited by 
$(\lambda,\rho)=(1.0,0.6)$ and $(\lambda,\rho)=(0.8,1.0)$: in the worst case, the total uncovered exposure is 18\% bigger than the total default fund. 
These values of $\lambda$ and $\rho$ correspond, however, to very intense and rather unrealistic credit and liquidity shocks, for which CMs lose the whole amount 
of a loan to a defaulted counterpart, and have no other means to get liquidity than fire selling their assets.

Results for the {\em cover 2} scenario (Figure \ref{fig5}) are different only for early reverberation rounds. 
Indeed for $n=2$ the default fund---already halved by covering the two most exposed CM, almost dries out for much lower values of $\lambda$ and $\rho$ as compared to the distributed shock scenario. 
Yet, total depletion of the default fund is observed again only for unrealistically high $\lambda$ and $\rho$, and for late shock propagation steps ($n>2$). 
Overall, we can again conclude that CC\&G default fund is robust in a wide range of model parameters (\ie, of economic scenarios), 
provided it is more conservatively gauged than prescribed by the {\em cover 2} requirement.

\section{Conclusions and future developments}\label{conclusion}

In this work we propose a new stress test methodology for central counterparties (CCPs) aimed at assessing the {\em vulnerability} (or equity at risk) of their clearing members (CMs). 
The model is based on a network characterization of CMs, whose balance sheets represent their financial situation and a CM is considered solvent as long as its equity is positive. 
In order to calculate daily financial position of each CM, we use a Merton-like model whose input is publicly available information. 
CMs are linked to each other through direct interbank credits and debits, that constitute the ground for the propagation of financial distress.
An initial shock is applied to the system to reduce the equity of each CM. The shock is made up of two components: an {\em exogenous} component, with a stochastic Poissonian shock 
and a deterministic shock, and an {\em endogenous} component, represented by an increase in margins to be posted to the CCP. 
The dynamics of financial distress propagation then combines two contagion channels: credit and liquidity shocks. 
Credit losses are related to counterparty risk and are faced by lender CMs when their borrower CMs get closer to default and struggle to fulfill their obligations. 
These losses can thus affect lenders, resulting in another wave of credit shocks. Liquidity shocks concern CMs that are unable to replace all the liquidity previously granted to them 
so they start fire selling their assets, which implies effective losses as illiquid assets trade at a discount. These shocks then reverberate throughout the market, turning into equity losses for other CMs. 
Note that in our model fire sales happen because banks are in need to recover lost fundings. However, there are other ways fire sales can originate from, 
such as correlated sales spirals due to common assets holdings, the leverage targeting policy adopted by banks~\citep{Greenwood2015,Duarte2015,Battiston2015X}
or liquidity hoarding behavior by CMs, which can further exacerbate the effect of liquidity shocks.

We remark that the stress propagation model builds on the assumption that equity losses experienced by a CM do imply both a decreasing value of its obligations
and a decreasing ability to lend money to the market---even if no default has occurred. The model then assesses {\em potential} losses for CMs resulting from a virtual dynamics of shocks propagation 
and thus, in a conservative way, does not include the explicit possibility for CMs to rearrange their balance sheet positions. 
However, we do model exogenous effects: a loss given default $\lambda<1$ corresponds to CMs being able to recover part of their loans to distressed institutions, 
whereas, a lost funding recovery $(1-\rho)>0$ allows banks to replace a fraction of lost liquidity with own cash reserves or from central banks before liquidating assets. 
Additionally, we model a potential external regulatory intervention by halting the contagion process at early propagation steps. 

The model here described constitutes an advancement in the existing stress testing methodologies and responds to ESMA's call for an innovative modeling of interconnections in the financial system.
We have applied this methodology to the Fixed Income asset class of CC\&G, the Italy-based clearing house, whose main cleared securities are Italian Government Bonds.
However, the model can be easily extended to cover multiple asset classes within the same CCP, as well as CMs belonging to multiple CCPs. 
This would allow to build a fully comprehensive stress testing framework, that considers the whole financial landscape as a single system made up of interconnected CCPs and CMs.

Numerical results obtained in this paper show that CC\&G's default fund, which is generally gauged on a {\em cover 4} basis, is adequate to cover losses recorded in the system after the introduction 
of a diversified set of initial shocks and their propagation in the network: even after an unlimited reverberation of distress within the system, the default fund is still able to cover losses stemming from all the defaulted CMs.
We also test the {\em cover 2} requirement by supposing that the two CMs to which CC\&G has the largest exposure under ``extreme but plausible'' market conditions default simultaneously and spread their distress over the network.
We observe that if we let shocks propagate unlimitedly (\ie, we suppose that no authorities intervention take place), the systemic impact of a {\em cover 2} initial shock is very similar to the one obtained 
with a distributed initial shock. However the {\em cover 2} initial shock produces a more severe impact on the system for early reverberations of financial distress, with higher CMs vulnerabilities and a bigger number of defaults. 
This result suggests that gauging the default fund on a {\em cover 2} basis might not be conservative enough, as additional defaults could be triggered. 
Indeed, measuring the effective losses a defaulting CM causes through its complex patterns of financial interconnections to the overall market secured by a CCP 
can lead to a more efficient definition of the default fund, as well as to a fairer default fund amount asked by the CCP to its CMs.

\section*{Acknowledgements}
This work was supported by the EU projects GROWTHCOM (FP7-ICT, grant n. 611272), MULTIPLEX (FP7-ICT, grant n. 317532), DOLFINS (H2020-EU.1.2.2., grant n. 640772) and the Italian PNR project CRISIS-Lab. 
The funders had no role in study design, data collection and analysis, decision to publish, or preparation of the manuscript.
This work was possible thanks to the support provided by Paolo Cittadini and colleagues of Cassa di Compensazione e Garanzia (CC\&G).

\bibliographystyle{apalike}

\end{document}